\documentclass[runningheads]{llncs}

\usepackage{graphicx}
\usepackage{tikz}
\usepackage{comment}
\usepackage{amsmath,amssymb} 
\usepackage{color}

\newcommand{\mat}[1]{\mathbf{#1}}

\newcommand{\norm}[1]{\left\lVert#1\right\rVert}

\begin{document}
\pagestyle{headings}
\mainmatter
\def\ECCVSubNumber{100}  

\title{From Perspective X-ray Imaging to Parallax-Robust Orthographic Stitching} 


\titlerunning{Abbreviated paper title}
%

\author{Javad Fotouhi\inst{1} \and
Xingtong Liu\inst{1} \and
Mehran Armand\inst{1} \and \\
Nassir Navab\inst{1, 2} \and
Mathias Unberath\inst{1}}

\authorrunning{J. Fotouhi et al.}
%
\institute{Johns Hopkins University, USA \and
Technical University of Munich, Germany \\ 
\url{\footnotesize{J. Fotouhi and X. Liu are joint first authors}} \\
\email{javad.fotouhi@jhu.edu}}
\maketitle

\begin{abstract}
Stitching images acquired under perspective projective geometry is a relevant topic in computer vision with multiple applications ranging from smartphone panoramas to the construction of digital maps. Image stitching is an equally prominent challenge in medical imaging, where the limited field-of-view captured by single images prohibits holistic analysis of patient anatomy. The barrier that prevents straight-forward mosaicing  of 2D images is depth mismatch due to parallax. In this work, we leverage the Fourier slice theorem to aggregate information from multiple transmission images in parallax-free domains using fundamental principles of X-ray image formation. The semantics of the stitched image are restored using a novel deep learning strategy that exploits similarity measures designed around frequency, as well as dense and sparse spatial image content. Our pipeline, not only stitches images, but also provides orthographic reconstruction that enables metric measurements of clinically relevant quantities directly on the 2D image plane.




\keywords{X-ray, Stitching, ConvNet, Landmark, GAN, and Parallax}
\end{abstract}

\section{Introduction}

\begin{figure}
  \centering
  \includegraphics[width=\textwidth]{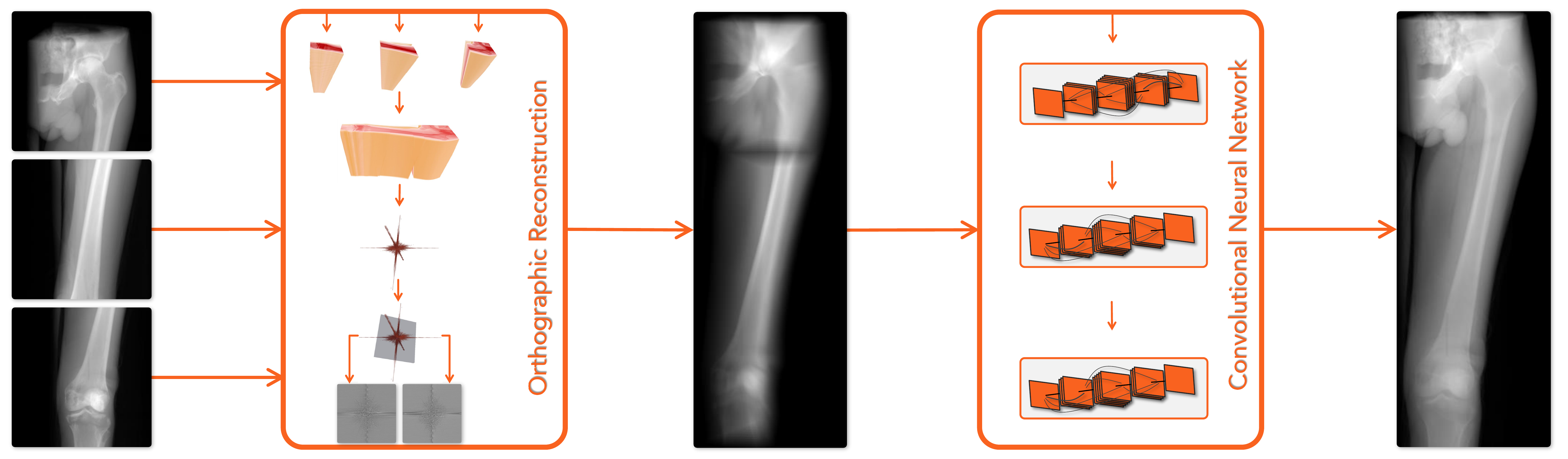}
  \caption{The image stitching pipeline includes orthographic 2D reconstruction of multiple 2D acquisitions, followed by restoration of image semantics using ConvNets.}
    \label{fig:1}
\end{figure}


Any two images of a planar scene are related by a homography $\mat{H} \in \mathbb{R}^{3 \times 3}$, an invertible mapping. Since the homography expresses the warping of the planar image content, these images can easily be stitched. In the special case where the camera motion is limited to rotation about its origin, homography can also describe the relative mapping as if the scene was at infinity. In all other cases, however, if $\mat{H}$ is used in an attempt to stitch images, it creates ghosting effects, also known as parallax. This has practical implications for every application where multiple images are acquired to capture an extended view of an arbitrary scene. In this case, the motion of the camera would typically comprise of both rotational and translational components. Unfortunately, and as the result of camera translation, parallax occurs for objects at different depths. The parallax effect appears stronger for points closer to the camera, and weaker for points farther from the camera. Due to parallax and inconsistencies between views, no universal mapping exists for the stitching of 2D images. 









\subsection{Related Work}

The problem of stitching has been long investigated in computer vision for various applications, including scene rendering and video frame stitching~\cite{sawhney1998robust,capel1998automated}. In its general form, since the mapping between images cannot be explained by a global homography, multiple early works addressed stitching by using local alignment strategies. In cases of minor translations with weak motion parallax, these works were able to deliver seamless mosaics by employing deghosting models based on local alignment within small patches in the overlapping region~\cite{szeliski1997creating,shum1998construction,szeliski2007image}. Blurry areas within the mosaics were reduced by limiting the sampling to the content of a single image per region, selected based on geometric information from prior segmentation~\cite{davis1998mosaics}. To construct visually appealing renderings, smooth transitions of features were suggested based on different depth cues~\cite{zhi2011toward}. Later works introduced hybrid stitching schemes by combining global homography and content-preserving local warping to render natural-looking mosaics~\cite{zhang2014parallax,lin2015adaptive}. Considering that all the aforementioned works failed to address the stitching problem in the presence of large motion parallax, an iterative stitching method was proposed to identify reliable local alignments and address depth mismatch with larger parallax~\cite{lin2016seagull}. Despite improving the warping between images, this work still suffered from poor stitching when parallax took place near the periphery of common regions.


In a clinical context, a common approach to stitching fluoroscopic images is to employ radiopaque planar markers that are placed parallel to the patient, approximately at the same depth from the camera as the anatomy~\cite{yaniv2004long,messmer2006image,chen2015ruler}. For instance, Yaniv et al. stitched X-ray images using a radiopaque ruler. They considered a simplified motion for the X-ray camera, restricted to fronto-parallel acquisitions, such that the planar mapping between images was parameterized using only two \textit{in-plane} translation and one \textit{in-plane} rotation parameters, hence ignoring all \textit{out-of-plane} variables. Their stitching pipeline was followed by a parallax correction step that estimated parallax errors via approximating reconstruction planes for each individual pixel. In a different work, Wang et al. proposed a parallax-free X-ray image stitching methodology by restricting the X-ray camera to only undergo pure rotation~\cite{wang2009parallax,wang2010parallax}. To compensate for translation between the C-arm camera and the scene, the patient bed was translated with the same translational parameters as the X-ray source. The movement of patient bed in the surgery room is not practical in most surgical environments. Instead, this solution can be delivered by a robotic C-arm platform that uses kinematics to enforce the rotational motion. Since robotic platforms typically cost an order of magnitude more than non-robotic scanners, from an economic standpoint, their availability for standard operative procedures will remain a concern. 

With the availability of 3D intra-operative imaging, instead of 2D X-ray image stitching, recent works suggested stitching of tomographic volumes reconstructed from X-ray images acquired on a circular trajectory~\cite{fuerst2015vision,fotouhi2018automatic}. While volumetric data do not suffer from the effects of parallax and can, therefore, be stitched easily, circular shortscans required for 3D tomographic reconstruction expose the patient to much higher radiation doses compared to simple X-ray acquisitions, suggesting that these methods are best employed only once, e.g., for verification.

\subsection{Clinical Motivation}
C-arm imaging offers fluoroscopic capabilities and is widely accepted as the workhorse imaging modality to guide minimally-invasive interventions across diverse specialities including orthopedics, neuro-, and endovascular surgery. Despite its advantage in providing anatomy level imaging, it suffers from limited field-of-view (FOV), resulting in truncated images of large structures. Prominent examples of such challenging cases are the appropriate reconstruction of comminuted long bone fractures or spinal fusion in scoliosis treatment. From a hardware standpoint, capturing the full extent of the anatomy requires imaging detectors at least as large as the target anatomy, which considering the ergonomics and workflow of surgery, is impractical.  

In fracture care surgery, it is of utmost importance to achieve proper alignment between the ends of extremities and reduce discrepancies to the contralateral side. It is commonly acknowledged that the verification mechanisms from 2D images are mentally challenging and cumbersome, an unfortunate circumstance that may lead to malrotation ($>15^{\circ}$) in approximately $30\%$ of femoral nailing cases \cite{ricci2001angular,jaarsma2004rotational}. These errors result in limping, pain, impaired walking, and may often require correction surgeries. The unmet clinical needs for intra-operative measurement of different biomechanical axes and bone lengths are also important concerns in other fields of orthopedic surgery, including high tibial osteotomy and total hip arthroplasty~\cite{hankemeier2006navigated,hofmann2008minimizing,plaass2011influence}.

Compositions of X-ray images were investigated to assist with quantifying the total length and angular measurements in pre- or post-operative settings for patients that underwent osteotomy, endoprosthesis, or fracture reduction procedures~\cite{hamer2004amorphous,boewer2005length}. These works do not address the image stitching problem and instead focus on disambiguating relative poses between multiple 2D images by using radiopaque scales that are placed approximately parallel to the extremity. Other approaches use bi-planar X-ray scanners with orthogonal planes and recover 3D anatomical landmarks through stereo 3D reconstruction~\cite{guenoun2012reliability}.

\subsection{Proposed Solution}
We suggest an end-to-end solution that combines information from 2D X-ray acquisitions in parallax-free domains, i.e., in 3D spatial and 3D Fourier spaces, and provides stitching of X-ray images with no constraints on the motion of the X-ray camera. Our mapping for stitching is no longer a homography; instead, it directly uses the projection matrices. The global structure of the orthographically stitched image is recovered by leveraging the Fourier slice theorem and principles governing image formation. The details in the stitched image are restored via a convolutional neural network (ConvNet) with regularization losses on frequency, as well as sparse and dense spatial features. 

As the result of employing a data-driven approach for learning the stitching parametrization and performing image-based rendering, the model we present is anatomy-specific, therefore, not directly comparable to the previous work that was invariant to the content of the image but was constrained by the motion of the camera~\cite{wang2010parallax}. To this end, we only focused on images of healthy human femurs in this work. 

Our proposed methodology \textit{i)} is invariant to parallax, \textit{ii)} provides orthographic reconstruction which enables direct metric measurements on the image without using any priors, \textit{iii)} is robust to minor gaps between input images such that missing information is recovered based on structure continuity learned by the ConvNet, and \textit{iv)} does not require explicit blending of content between multiple sources in their overlapping regions.


\section{Methodology}
Our solution to transmission image stitching shown in Fig.~\ref{fig:1} is designed based on two fundamental steps. First, we leverage the close relationship between the Fourier slice theorem and the Radon transform to provide a stitched image in an orthographic geometry from back-projected rays (Sec.~\ref{subsec:ortho_recon}). Second, to restore the missing and blurred content, we use a series of ConvNets with adversarial losses and regularizes for structural, intensity, contrast, frequency, and sparse feature similarities between input and ground-truth orthographic images (Sec.~\ref{subsec:convNet}). 

We simultaneously train an anatomical landmark detector network, which serves two purposes: \textit{i)} it automatically detects anatomical landmarks that are critical to metric and angular measurements from the bone on the stitched image, and \textit{ii)} it integrates into the stitching pipeline and enforces the network to predict images closer to the ground-truth domain such that the landmark detection with an identical network performs well on both the ground-truth and prediction domains.

\subsection{Orthographic Reconstruction}\label{subsec:ortho_recon}
The back-projection of each pixel element $(x_m, y_n)$ in a 2D X-ray image $g(x_m, y_n)$ is defined as:
\begin{equation}
    V(\mu(d, i, m, n)) = P_i^+ g(x_m, y_n)\quad \text{, where}
\end{equation}
$\mu(d, i, m, n)$ is the ray characterized by the projection matrix $P_i$ and pixel coordinates $(x_m, y_n)$ in the $i$-th image, and $V(.)$ is the volume constructed by smearing out $\mu$ into the 3D space. The parameter $d \in [0, 1]$ is used identically for all projections. It defines the depth of the backprojected volume, where $1$ refers to backprojection within the entire imaging cone, and $0.5$ denotes backprojection between the detector plane and the depth equivalent of $50\%$ of the focal length. The $(.)^+$ denotes the pseudo-inverse operation. Since all projection images are acquired from an identical static scene, the back-projected rays from all acquisitions can be compounded into a single volume as:
\begin{equation}
    \Omega = \sum_{i, m, n} V(\mu(d, i, m, n)). 
\end{equation}

\begin{figure}
  \centering
  \includegraphics[width=\textwidth]{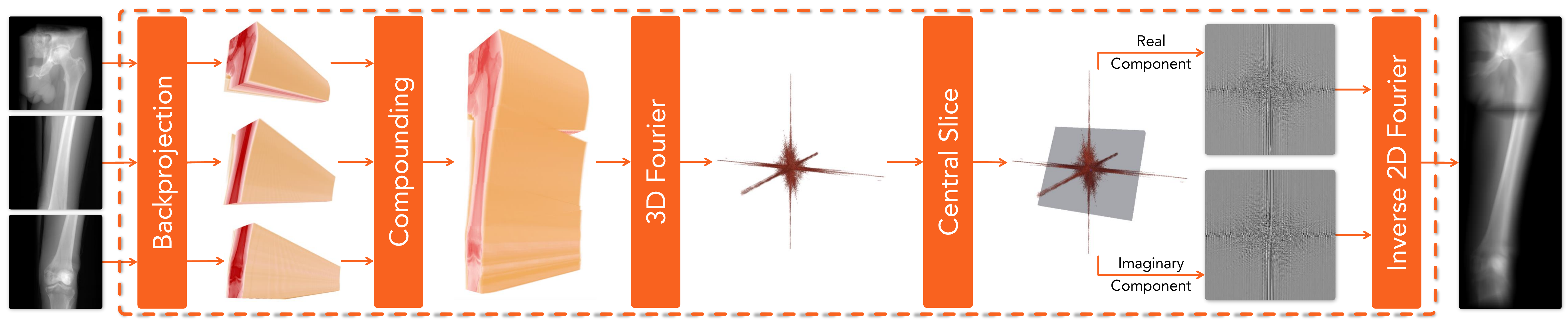}
  \caption{The orthographic reconstruction pipeline leverages the Fourier slice theorem to transform multiple 2D X-ray images acquired using cone-beam geometry, to a single extended-view image in parallel-beam geometry. In this example, the backprojection is performed with $d=0.5$, hence in a volume between the detector plane and $50\%$ of the focal length.}
    \label{fig:2}
\end{figure}

In the remainder of this section, we treat the stitching of multiple input images, as a reconstruction problem that aims to reconstruct an orthographic 2D view given the incomplete 3D data $\Omega$. As shown in Fig.~\ref{fig:2}, within the orthographic image, the global structure of the scene is reconstructed with insufficient details due to \textit{i)} loss of information in $\Omega$ as it is only constructed from sparse and single-view set of data, and \textit{ii)} different binning and sampling of frequencies, particularly in the overlapping region, which can be explained by the Fourier-slice theorem~\cite{turbell2001cone}.

Based on the Fourier slice theorem, also known as the projection slice theorem, the Fourier transform of an orthographic projection of a 3D function $f(x, y, z)$ in 2D represented as $G(\Theta)$, is equivalent to the 2D slice in the 3D Fourier of the function $f(x, y, z)$ that passes through the origin and is parallel to the projection plane $\Theta$. Given the Fourier transform $F = \mathcal{F}(\Omega)$, we represent the central slice that passes through the origin and is parallel to the projection plane $\Theta_{0}$ as $G(\Theta_{0}) = F(\theta = \Theta_{0})$. Based on the Fourier slice theorem, the stitched image $I$ in Fig.~\ref{fig:2} is then computed as $I = \mathcal{F}^{-1} (G(\Theta_{0}))$.


\subsection{Restoration of Stitching Semantics}\label{subsec:convNet}
In CT reconstruction, if X-ray images are acquired on a circular trajectory, oversampling occurs at the center of the Fourier domain. For the stitching problem in hand, since there is no analytical approach to identifying the appropriate filter for re-sampling and re-binning, as well as the loss of information from orthographic rendering of a perspective image, we suggest a learning strategy to jointly learn corrections in the spatial and frequency domains.

\begin{figure}[t]
  \centering
  \includegraphics[width=\textwidth]{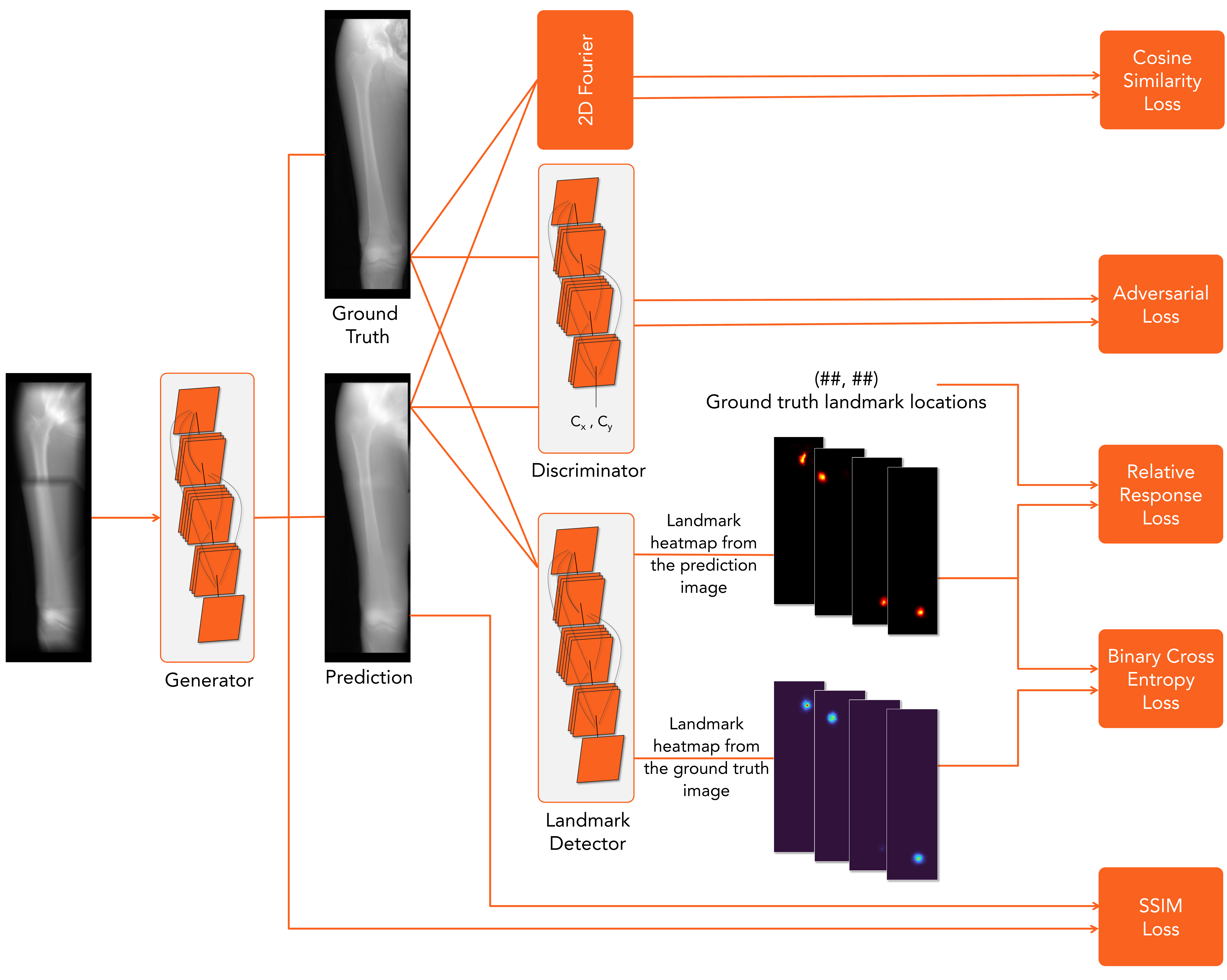}
  \caption{The overall network architecture that was used for training is shown. The input orthographic 2D reconstruction in the frequency domain is generated using our proposed method from several X-ray images. SSIM and adversarial losses are used to enforce the network to generate visually similar images compared to the ground-truth. Cosine similarity loss is employed to emphasize more on the high-frequency components of the predictions. RR and BCE loss are used to train a landmark detection network and encourage the semantics restoration to be functionally similar to the ground-truth reconstruction.}
    \label{fig:network}
\end{figure}

The learning framework, as shown in Fig.~\ref{fig:network}, consists of three modules that are trained in an end-to-end fashion. The first module, also known as the generator, restores image semantics from the 2D reconstructions introduced in Sec.~\ref{subsec:ortho_recon}. To better represent the underlying structural information of the anatomy, the semantics are produced as continuous values instead of discrete categories. The second module is the discriminator that attempts to distinguish between the predicted images and the ground-truth 2D orthographic projections. This encourages the generator module to predict visually similar images compared with the ground-truth domain. The last module detects anatomical landmarks from the orthographic 2D reconstructions. This module not only facilitates automatic landmark detection but also encourages the predicted images to offer more details near the important structures of the anatomy.

Structural similarity (SSIM) loss~\cite{wang2004image} and an adversarial loss on the spatial domain as well as a cosine similarity loss on the frequency domain are used to optimize the overall visual similarity between the predicted images and the ground-truth orthographic projections. The SSIM index comprises a weighted multiplication between three distance measurements, namely luminance $l$, contrast $c$, and structure $s$, between the prediction $X$ and ground-truth projection $Y$. The loss between $X$ and $Y$ is defined as:
\begin{equation}
\mathcal{L}_{\text{ssim}} = 1 - \frac{1}{\lvert\mathrm{\Omega}\rvert}\sum_{{\left(i,j\right)}\in\mathrm{\Omega}}l_{i,j}^{\alpha}c_{i,j}^{\beta}s_{i,j}^{\gamma}\quad \text{, where}
\end{equation}
$i,j$ iterate over the entire image domain. The three terms in SSIM are computed as $l_{i,j}=\frac{2\mu_{i,j}^x\mu_{i,j}^y+b_1}{{\mu_{i,j}^x}^2+{\mu_{i,j}^y}^2+b_1}$, $c_{i,j}=\frac{2\sigma_{i,j}^x\sigma_{i,j}^y+b_2}{{\sigma_{i,j}^x}^2+{\sigma_{i,j}^y}^2+b_2}$, and $s_{i,j}=\frac{\sigma_{i,j}^{xy} +b_3}{\sigma_{i,j}^x\sigma_{i,j}^y+b_3}$, such that $\mu_{i,j}^x$ and $\sigma_{i,j}^x$ are the local window average and standard deviation of $X$ centered at location $\left(i,j\right)$, respectively. $\sigma_{i,j}^{xy}$ is the local window covariance of $X$ and $Y$ centered at location $\left(i,j\right)$.

The adversarial loss follows the basic idea of relativistic GAN~\cite{jolicoeur2018relativistic}. The adversarial losses in the discriminator and generator cycles are:
\begin{equation}
\begin{aligned}
&\mathcal{L}_\text{D} = \left(1 - C_y + \overline{C_x}\right)^2 + \left(1 - \overline{C_y} + C_x\right)^2 \\
&\mathcal{L}_\text{G} = \left(1 - C_x + \overline{C_y}\right)^2 + \left(1 - \overline{C_x} + C_y\right)^2\quad\text{, where}
\end{aligned}
\end{equation}
$C_x$ is the confidence of the discriminator regarding whether the prediction $X$ is real and $\overline{C_x}$ is the average confidence over a mini-batch. The architecture of the discriminator is DenseNet~\cite{jegou2017one} with linear activation as the final layer. 

In the frequency domain, the proposed cosine similarity loss is defined as:
\begin{equation}
    \mathcal{L}_\text{cos} = 1 - \langle\frac{f_x - f_i}{\norm{f_x - f_i}_2}, \frac{f_y - f_i}{\norm{f_y - f_i}_2}\rangle\quad \text{, where}
\end{equation}
$f_x$ is the frequency representation of the restored semantics $X$ that is flattened to a 1D vector, and $f_i$ is the frequency representation of the input orthographic reconstruction. Since the low-frequency components in images are much larger than the high-frequency ones, we only use the residual frequency for both the prediction and the ground-truth reconstruction w.r.t. the input reconstruction. This could encourage the network to focus more on the high-frequency texture information in the ground-truth.

To adequately convey task-related information to the generator network (semantic restoration network), we propose to jointly optimize the generator network as well as an anatomical landmark detection network with a specialized training scheme. The landmark detector model takes either ground-truth reconstruction or the predictions of the generator as input and predicts heatmaps corresponding to distinct anatomical landmarks. The ground-truth landmark locations and heatmaps are generated and annotated from the CT volume and used for supervision. Each ground-truth heatmap is a 2D Gaussian distribution with a corresponding landmark location as the mean and a manually selected standard deviation $\sigma$. Binary Cross Entropy (BCE) and Relative Response (RR) losses~\cite{liu2020extremely} are used for the landmark detection learning. The BCE loss is defined as follows:
\begin{equation}
    \mathcal{L}_\text{bce} = \frac{1}{\lvert\mathrm{\Omega}\rvert}\sum_{{\left(i,j\right)}\in\mathrm{\Omega}}m^\text{gt}_{i,j}\log{m_{i,j}} + \left(1 - m^\text{gt}_{i,j}\right)\log\left(1 - m_{i,j}\right)\quad\text{, where}
\end{equation}
$m_{i,j}$ and $m^\text{gt}_{i,j}$ are the values of the predicted and ground-truth heatmaps, respectively, at the location $\left(i,j\right)$. RR loss is defined as: 
\begin{equation}
\mathcal{L}_\text{rr} = -\log\left(\frac{\text{e}^{\sigma m_{u,v}}}{\sum_{{\left(i,j\right)}\in\mathrm{\Omega}}\text{e}^{\sigma m_{i,j}}}\right) \quad \text{, where}
\end{equation}
$\left(u,v\right)$ is the ground-truth landmark location and $\sigma$ is a scale factor. 

In a single iteration of training, there are several cycles involved, where only one of all modules are updated. In the cycle of discriminator training, only $\mathcal{L}_\text{D}$ is involved. In the cycle of landmark detector training, the overall loss is $\mathcal{L}_\text{landmark} = \lambda_\text{rr}\mathcal{L}_\text{rr} + \lambda_\text{bce}\mathcal{L}_\text{bce}$, where the input to the landmark detector is the ground-truth 2D orthographic projections. In the cycle of semantics restoration training, the overall loss is $\mathcal{L}_\text{restore} = \lambda_\text{ssim}\mathcal{L}_\text{ssim} + \lambda_\text{G}\mathcal{L}_\text{G} + \lambda_\text{cos}\mathcal{L}_\text{cos} + \lambda_\text{rr}\mathcal{L}_\text{rr} + \lambda_\text{bce}\mathcal{L}_\text{bce}$, where the input to the detector is the predicted stitching images. The landmark detector is only updated when the ground-truth reconstruction is fed to ensure that it only learns features that appear in the ground-truth reconstruction. When the stitching predictions are used as input to the landmark model, only the generator network gets updated, which implicitly forces the stitching generator network to learn spatial features in the ground-truth domain that contribute to the task of landmark detection.

\section{Experiments and Results}

\subsection{Data Set} 
The training data comprises of eight CT volumes from the left and right legs of four cadaveric specimens. We generated the training data by simulating realistic digitally-reconstructed radiographs (DRRs) using the physics-based DeepDRR pipeline~\cite{unberath2018deepdrr,unberath2019enabling}. Rather than relying on full Monte Carlo simulation of image formation, DeepDRR  analytically generates forward projections from CT volumes by accounting for the physical interactions that occur during image formation and then estimates the contributions of scattering and noise. Compared to naive DRRs, this mechanism has demonstrated improved generalizability~\cite{bier2018x,bier2019learning}.
\begin{figure}
  \centering
  \includegraphics[width=\textwidth]{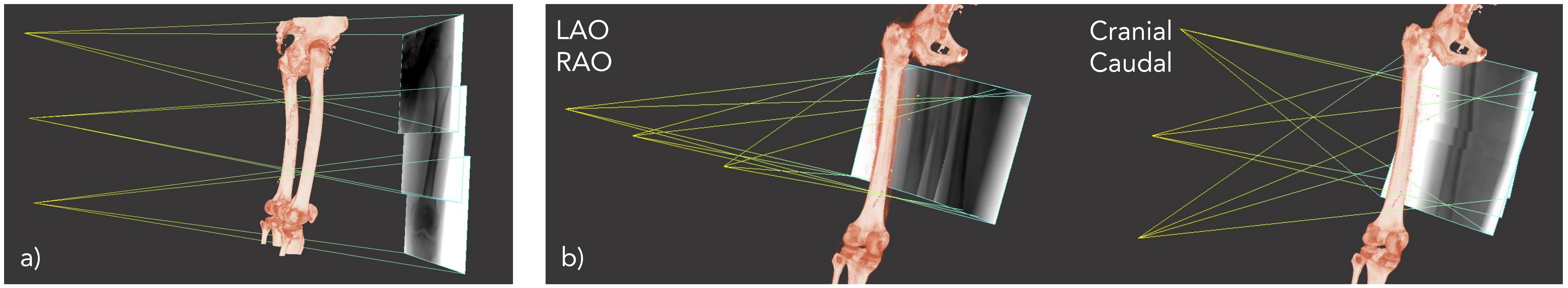}
  \caption{\textbf{a)} Each training instance comprises three X-ray images. \textbf{b)} The training X-ray images are generated in the LAO/RAO and Cranial/Caudal directions.}
    \label{fig:3}
\end{figure}

The intrinsic parameters of the X-ray camera are selected based on the nominal parameters of a commercially available flat panel C-arm, Cios Fusion (Siemens Healthineers, Forchheim, Germany). We generated a total of $45,000$ X-ray images, where each $3$ were stitched together, resulting in $15,000$ training instances, as shown in Fig.~\ref{fig:3}-a. The translation components of the X-ray camera were adjusted such that the first image is acquired from the region of the femoral head, the second image from the shaft, and the third from the knee. We also added a random value in the range of $[-20, 20]$\,mm to each translation component in the $x, y, z$ axes, which resulted in gaps or overlaps between simulated acquisition. 

For each specimen's leg, we generated DeepDRR images from $-21^{\circ}$ to $+21^{\circ}$, and $-6^{\circ}$ to $+6^{\circ}$ in LAO/RAO and Cranial/Caudal directions, respectively, around the anterior-posterior view of the bone. These directions which correspond to \textit{out-of-plane} rotations, are shown in Fig.~\ref{fig:3}-b. We also added an offset up to $[-6^{\circ}, 6^{\circ}]$ to each \textit{out-of-plane} rotation element of the X-ray camera. This rotation offset is intended to prevent all three images from being exactly parallel to each other, and enforce the training to become robust to unintended \textit{out-of-plane} rotations. In the backprojection step, all projection images were smeared out with $d=0.5$, which realistically assumes the imaged object is at most $500$\,mm away from the detector.

A validation set is constructed from $1875$ DeepDRR images from a separate cadaveric CT. Finally, for testing, we generated a total of $3750$ images from the left and right CT scans of two other patients. 

In our supervised training scheme, the ground-truth images were obtained by generating forward projections of the original CT volumes in an orthographic model (parallel-beam geometry). Given the rotation matrix $R$ associated with the ray direction, the orthographic projection is defined as~\cite{hartley2003multiple}:
\begin{equation}
    \begin{bmatrix} 
        x \\ y \\ \gamma \end{bmatrix} = 
    \begin{bmatrix} 
        1 \: & 0 \: & 0 \: & 0 \: \\ 0 & 1 & 0 & 0 \\ 0 & 0 & 0 & 1 
    \end{bmatrix} 
    \begin{bmatrix} 
        R \: & \mathbf{0} \: \\ \mathbf{0}^{\top} & 1 
    \end{bmatrix} 
    \begin{bmatrix}
        X \\ Y \\ Z \\ 1
    \end{bmatrix}
    = 
    \begin{bmatrix}  
        \mathbf{r}_1^{\top} \: & 0 \\ \mathbf{r}_2^{\top} \: & 0 \\ \mathbf{0}^{\top} & 1 
    \end{bmatrix}
    \begin{bmatrix}
        X \\ Y \\ Z \\ 1
    \end{bmatrix} \quad \text{, where}
\end{equation}
$\mathbf{r}_i^{\top}$ is the $i$-th row in $R$, and $(\frac{x}{\gamma}, \frac{y}{\gamma})^{\top}$ are the 2D projections of the 3D voxels that are denoted by the homogeneous coordinates $(X, Y, Z, 1)^{\top}$. 


\subsection{Stitching Performance}
\begin{figure}
  \centering
  \includegraphics[width=\textwidth]{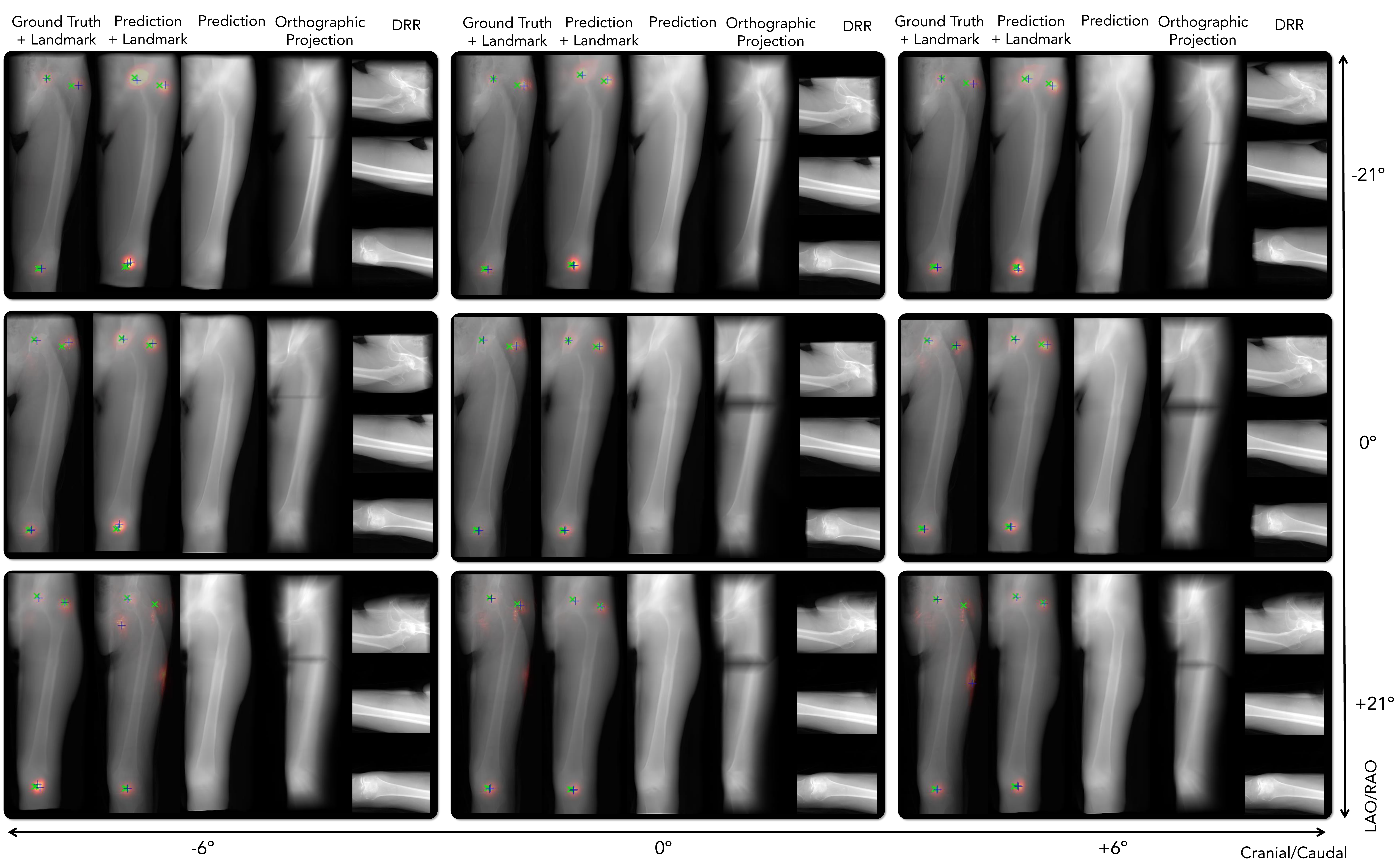}
  \caption{Stitching outcomes are shown on test data from the left leg. The heatmaps corresponding to different landmarks are shown as red overlays. The ground-truth and predicted landmark locations are shown as green and blue crosses, respectively.}
    \label{fig:4}
\end{figure}

\begin{figure}
  \centering
  \includegraphics[width=\textwidth]{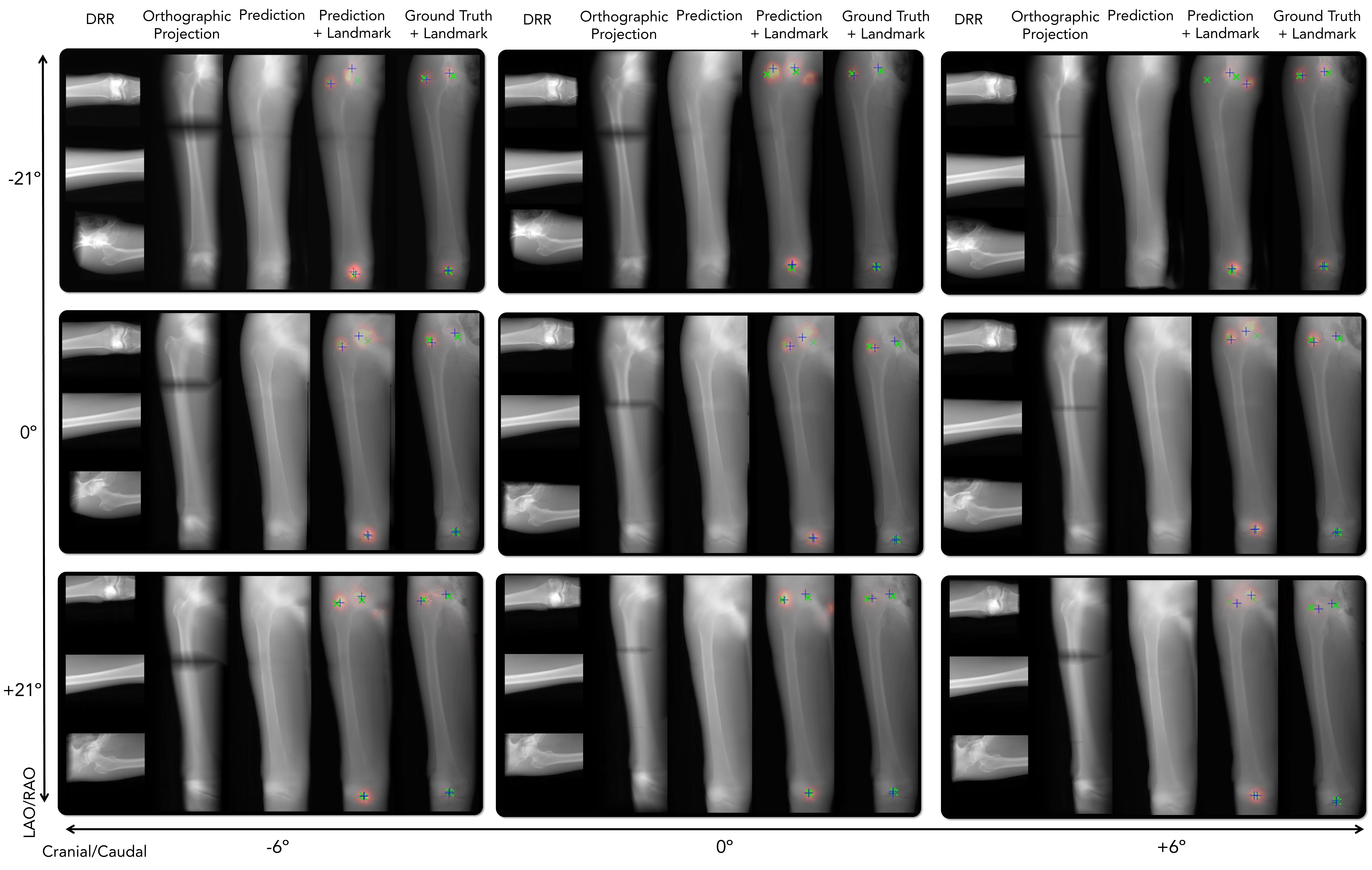}
  \caption{Stitching outcomes are shown on test data from the right leg. The heatmaps corresponding to different landmarks are shown as red overlays. The ground-truth and predicted landmark locations are shown as green and blue crosses, respectively.}
    \label{fig:5}
\end{figure}

The stitching results on two test data sets are shown in Figs.~\ref{fig:4} and~\ref{fig:5}. It should be noted that in all cases that \textit{out-of-plane} rotations were present between the three input data, the orthographic reconstruction plane was selected to be parallel to the first input image. In Figs.~\ref{fig:4} and~\ref{fig:5}, we also demonstrate the detection of four bony landmarks, namely the femoral head, greater trochanter, patellar groove (knee), and tibia. These landmarks are particularly important as they are used in defining the length and the biomechanical axes of the bone.

We trained for $58$ epochs on the data with $1$\,mm pixel spacing, and with the input and output image sizes of $640 \times 640$\,pixels. Using this model, we achieved an SSIM similarity score of $95.7\%$ and a PSNR of $25.70$\,(db). The BCE and RR landmark losses were $0.0044$ and $13.53$, respectively.

\begin{table}
\centering
\caption{Comparing the performance of the network given different regularization factors $\lambda_\text{cos}$ for the cosine frequency loss in an ablation study}
\label{tab:table_cosine}
\begin{tabular}{l|c|c|c|c|c|c}
 & $\lambda_\text{cos} = 0.0$ & $\lambda_\text{cos} = 1.0$ & $\lambda_\text{cos} = 2.0$ & $\lambda_\text{cos} = 3.0$ & $\lambda_\text{cos} = 4.0$ & $\lambda_\text{cos} = 5.0$ \\ \hline
\textbf{SSIM}\:($\%$) & $92.56$ & $94.66$ & $95.81$ & $\mathbf{95.87}$ & $95.27$ & $94.23$ \\
\textbf{PSNR}\:(db) & $24.85$ & $26.80$ & $\mathbf{26.83}$ & $25.58$ & $24.94$ & $24.83$ \\
\textbf{BCE} & $\mathbf{0.0045}$ & $0.0047$ & $0.0047$ & $0.0046$ & $0.0046$ & $0.0049$ \\
\textbf{RR} & $11.14$ & $10.75$ & $\mathbf{10.53}$ & $10.60$ & $10.79$ & $11.27$
\end{tabular}
\end{table}

\subsection{Ablation Study}
To identify the optimal learning parameters and understand the effects of each component, we performed ablation studies. To keep all experiments tractable, we completed the ablation tests on $15,000$ downsampled training data with $2$\,mm pixel spacing, with 10 epochs, and a batch size of 2. We kept the regularization factor for the GAN losses fixed and evaluated against all other regularization parameters.  

\begin{table}
\centering
\caption{Comparing the performance of the network given different regularization factors $\lambda_\text{rr}$ for landmark detection}
\label{tab:table_rr}
\begin{tabular}{l|c|c|c|c|c|c}
 & $\lambda_\text{rr} = 0.0$ & $\lambda_\text{rr} = 0.03$ & $\lambda_\text{rr} = 0.06$ & $\lambda_\text{rr} = 0.09$ & $\lambda_\text{rr} = 0.12$ & $\lambda_\text{rr} = 0.15$ \\ \hline
\textbf{SSIM}\:($\%$) & $95.31$ & $\mathbf{95.93}$ & $95.59$ & $94.72$ & $94.77$ & $94.47$ \\
\textbf{PSNR}\:(db) & $24.81$ & $\mathbf{26.16}$ & $25.97$ & $25.13$ & $25.06$ & $24.87$ \\
\textbf{BCE} & $0.0082$ & $\mathbf{0.0040}$ & $0.0042$ & $0.0047$ & $0.0060$ & $0.0071$ \\
\textbf{RR} & $26.23$ & $11.00$ & $10.57$ & $10.56$ & $\mathbf{10.41}$ & $10.74$
\end{tabular}
\end{table}

\begin{table}
\centering
\caption{Comparing the performance of the network given different regularization factors $\lambda_\text{bce}$ for landmark detection. N/A indicates that the training diverged with the corresponding parameters.}
\label{tab:table_sb}
\begin{tabular}{l|c|c|c|c|c|c}
 & $\lambda_\text{bce} = 0$ & $\lambda_\text{bce} = 100$ & $\lambda_\text{bce} = 200 $ & $\lambda_\text{bce} = 300$ & $\lambda_\text{bce} = 400$ & $\lambda_\text{bce} = 500$ \\ \hline
\textbf{SSIM}\:($\%$) & N/A & $95.17$ & $95.38$ & $95.54$ & $\mathbf{95.84}$ & $95.42$ \\
\textbf{PSNR}\:(db) & N/A & $25.19$ & $26.65$ & $23.63$ & $\mathbf{27.05}$ & $26.02$ \\
\textbf{BCE} & N/A & $0.0045$ & $0.0038$ & $0.0038$ & $\mathbf{0.0036}$ & $0.0039$ \\
\textbf{RR} & N/A & $11.01$ & $10.86$ & $10.93$ & $\mathbf{10.69}$ & $11.01$
\end{tabular}
\end{table}

In Tables~\ref{tab:table_cosine},~\ref{tab:table_rr}, and~\ref{tab:table_sb} we report the best losses by selecting the lowest validation loss. The tables report SSIM and peak signal-to-noise ratio (PSNR) similarity scores, as well as BCE and RR losses. The best results in each experiment are represented with bold digits.

\subsection{Landmark Detection}
\begin{figure}
  \centering
  \includegraphics[width=0.9\textwidth]{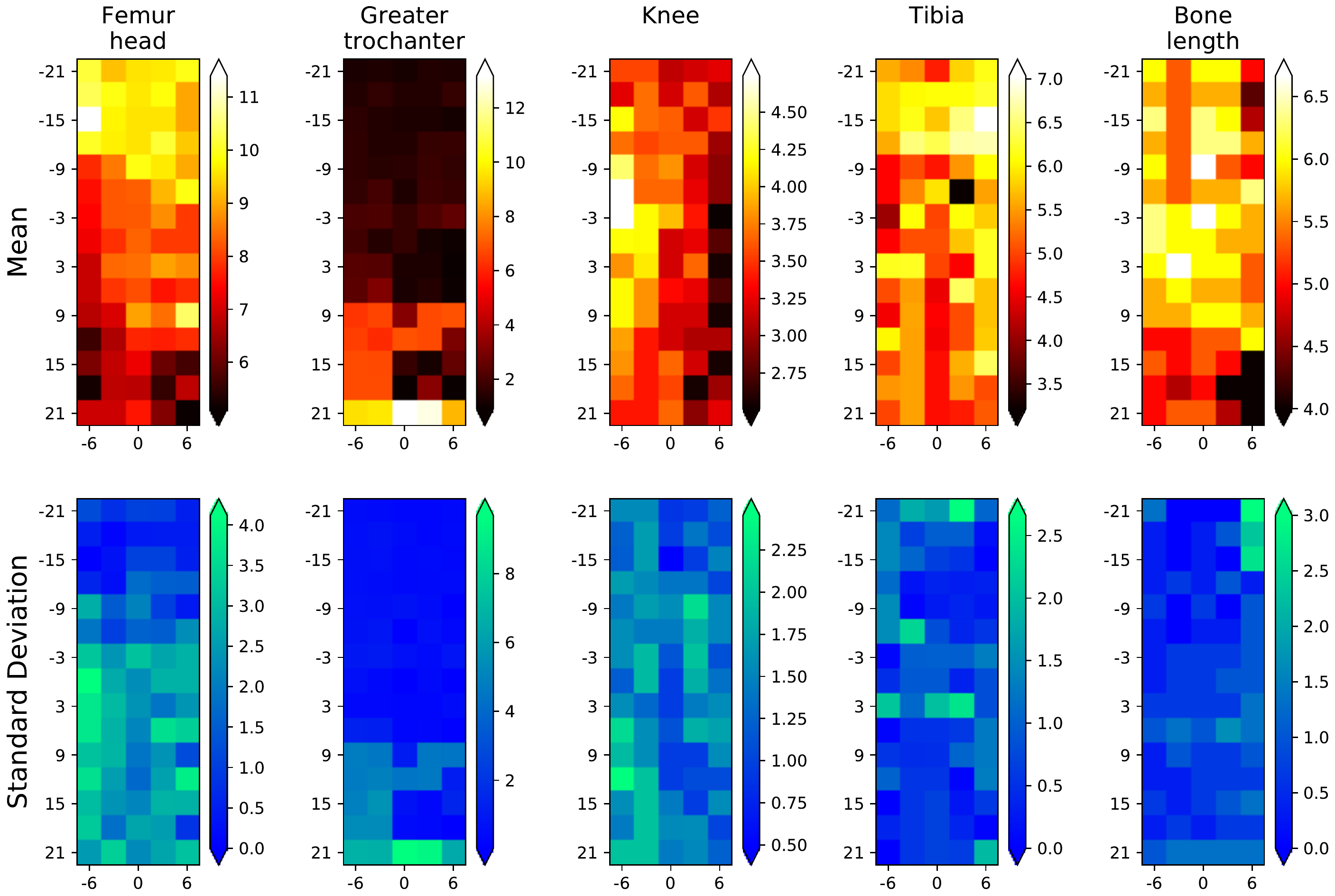}
  \caption{Landmark detection performances on the ground-truth orthographic projections for two test patient data sets are shown as error heatmaps. The vertical axes correspond to the LAO/RAO rotations, and the horizontal axes correspond to cranial/caudal rotations around the anterior-posterior view, respectively.}
    \label{fig:gt_landmark}
\end{figure}

\begin{figure}
  \centering
  \includegraphics[width=0.9\textwidth]{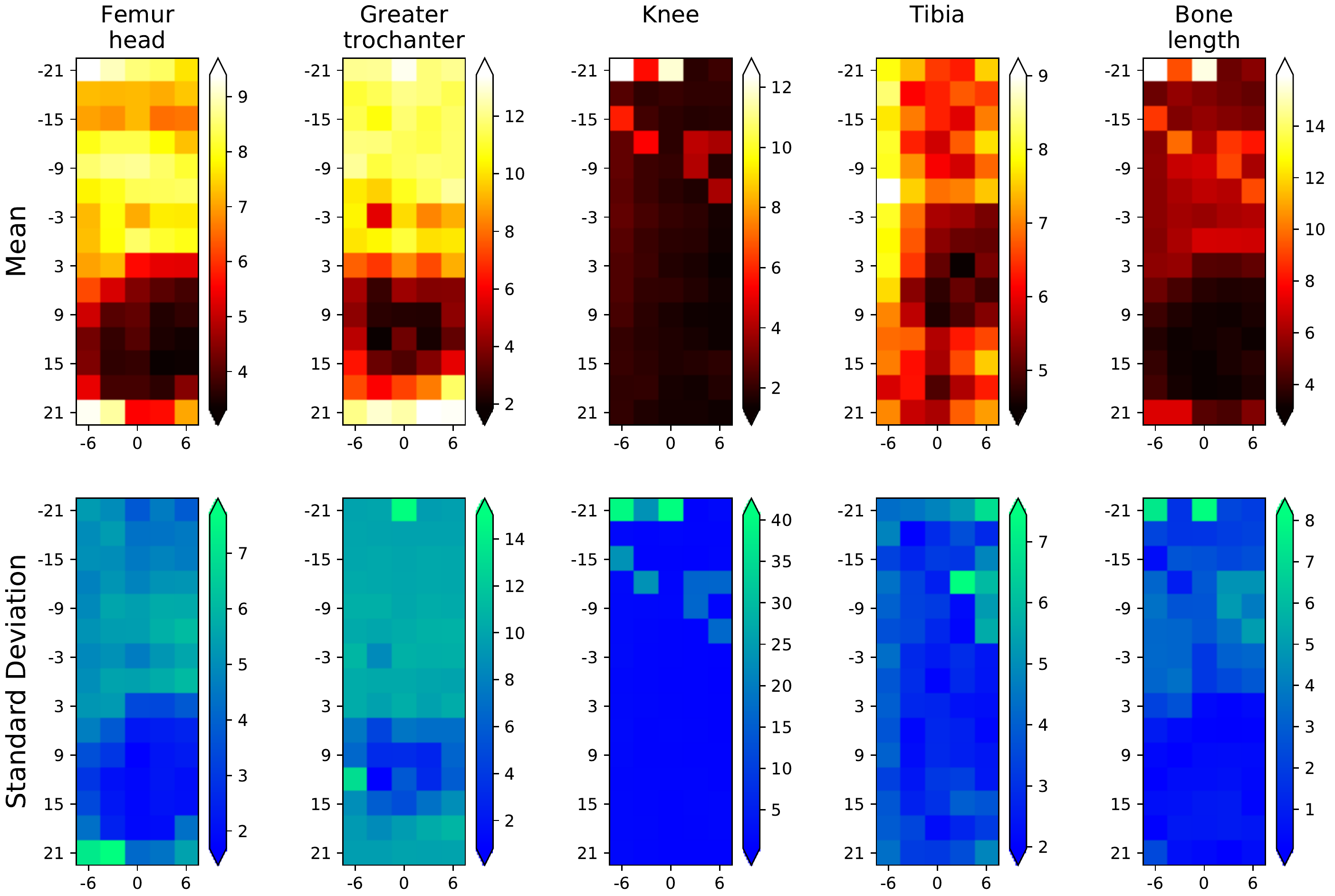}
  \caption{Landmark detection performances are shown on the prediction images of the test sets. The vertical axes correspond to the LAO/RAO rotations, and the horizontal axes correspond to cranial/caudal rotations around the anterior-posterior view, respectively.}
    \label{fig:prediction_landmark}
\end{figure}

We evaluated the accuracy of landmark detection on $3500$ test ground-truth and prediction images. The corresponding results are presented in Figs.~\ref{fig:gt_landmark} and~\ref{fig:prediction_landmark}. The rightmost columns in these figures contain the automatic measurements of total bone length directly from 2D orthographic ground-truth or prediction images.

\section{Discussion}
The outcome of orthographic stitching presented in Figs.~\ref{fig:4} and~\ref{fig:5} suggest that the ConvNet can effectively close small gaps between input images, and substantially reduce more significant gaps. The automatic landmark detection in Figs.~\ref{fig:gt_landmark} and~\ref{fig:prediction_landmark} indicate that the mean landmark detection errors are $5.06$ and $6.07$ pixels on the ground-truth projections and the prediction images, respectively. It is important to note that, despite the errors in automatic measurement of the bone length based on anatomical landmarks, the manual measurements from the prediction images were highly consistent with the measurements from the ground-truth orthographic projections. This suggests that the strict modeling of the geometry of stitching based on the projection matrices and the central slice theorem, and using the ConvNet to recover details, yielded true orthographic predictions.


We observed that object magnification that happens when the imaged object is closer to the origin of the X-ray camera has an adverse effect on orthographic reconstruction. When the object is closer to the origin, the rays from the source to the image plane that hit the anatomy are more diverged compared to the parallel rays that are used for reconstruction. For the reconstruction, we assumed the patient is within a $500$\,mm distance from the image detector, which is a clinically realistic assumption.

Lastly, in our ablation studies, we demonstrated that all regularization losses contribute to stronger image similarities for image reconstruction.

\section{Conclusion}
We presented the first work that provides orthographic image stitching by leveraging the principles of image formation and geometric models from computer vision and combining them with ConvNets to recover semantics. In contrast to the state-of-the-art systems, our solution naturally allows for both translation and rotations of the X-ray camera and does not impose any constraints regarding the motion. A direction for future research is to extend the training to complex scenarios where stitching becomes robust to fractures and arbitrary tools in the scene using robust estimators~\cite{esfandiari2018deep,gao2019localizing,belagiannis2015robust}.

The orthographic representations of images \textit{i)} do not carry perspective properties, hence enable metric measurements directly on 2D images, and \textit{ii)} are described with 5 \textit{degree-of-freedom} (DOF) as opposed to 6 DOF rigid body parameters. As a consequence of this drop of DOF, registration of orthographic images will not require scale disambiguation along their depth. 

In reflective images used in standard computer vision applications, the Fourier slice theorem does not hold. Nevertheless, other contributions of our work, such as back-projection, data compounding in parallax-free domains, and the recovery of details using ConvNets, can be directly employed to tackle image stitching or image-based rendering problems in other domains of computer vision applications.

\clearpage
%
%
\bibliographystyle{splncs04}

\end{document}